\documentclass[a4paper,11pt]{article}
\pdfoutput=1 % if your are submitting a pdflatex (i.e. if you have
\usepackage{jheppub} % for details on the use of the package, please
\usepackage[T1]{fontenc} % if needed

\title{\bf Displaced Vertices from Hidden Glue}

 \author{Gustavo  Burdman}
 \author{and Gabriela Lichtenstein}
 \affiliation{Department of Mathematical Physics,\\
Institute of Physics, \\
University of Sao Paulo,\\
R. do Matao 1371, Sao Paulo, \\
SP 05508-090, Brazil}

\emailAdd{burdman@fma.if.usp.br}
\emailAdd{gabriela.lichtenstein@usp.br}

\abstract{We consider the phenomenology of glueball production and
  decay in theories with hidden glue. We focus on the case of 
folded supersymmetry (FSUSY) , where there is an unbroken $SU(3)$ 
  gauge theory in the absence of light matter, leading to the formation
of glueballs. We study their production through the annihilation of
folded squarks into hidden gluons at the LHC, and  model their 
fragmentation  into glueballs. We obtain the
distribution of displaced vertices, which is directly determined by the folded
squark mass scale. Although here we specifically apply it to FSUSY,
the procedure to model the hidden glue fragmentation into glueballs
can be generalized to other similar theories.}

\begin{document} 
\maketitle
\flushbottom

\section{Introduction}
\label{sec:intro}

The discovery of the Higgs boson~\cite{Aad:2012tfa,Chatrchyan:2012xdj}
completes the spectrum of the
standard model (SM). It also confirms its validity at the weak scale,
where the gauge interactions have been tested with great precision,
and now the Higgs sector responsible for electroweak symmetry breaking
appears to be just as in the SM. It is clear that there are still many
questions left to answer. Among them, the origin of dark matter, the
large hierarchy of Yukawa couplings, the origin of neutrino masses and
of the matter--anti-matter asymmetry. However, the question that
potentially limits the validity of the SM the most is that of the
stability of the weak scale as defined by the Higgs sector. 
Since in the SM the Higgs potential is not protected by any symmetry,
it is highly sensitive to ultraviolet (UV) scales. This can be seen by
considering the loop contributions of various SM states to $m_h^2$,
resulting in quadratic sensitivity to the UV. In particular, since the
top quark is the state that couples the most to the Higgs boson its
contribution to $m_h^2$  determines the tuning of the Higgs mass
counterterm for a given value of the UV cutoff.   In order to avoid
large tunings the SM cutoff should not be too far above the TeV
scale. Thus, the absence of signals for physics beyond the SM
constitutes a problem: the SM must be highly tuned. 

Extensions of the
SM that address the Higgs sector UV sensitivity do so by introducing a
symmetry that protects the Higgs mass. For instance, in 
supersymmetric extensions~\cite{Martin:1997ns,Aitchison:2005cf}, super-partners of the SM states
-particularly of the top quark-  are
responsible to cancel the quadratic UV sensitivity of $m_h^2$.  These
are the (left and right-handed) scalar partners of the top quark, the
stops. On the other hand, in theories where the Higgs is a composite
state its mass is typically protected by a global symmetry, which is
spontaneously broken and has the Higgs as one of its
pseudo-Nambu-Goldstone bosons (pNGBs)~\cite{Bellazzini:2014yua,Panico:2015jxa}. In these theories, the symmetry partners of the
top quark are fermions that together with the top form representations
of the global symmetry group.  The fact that none of these new states
have been seen, either directly or through their indirect effects,
puts stringent bounds on their masses, pushing the SM
cutoff to higher scales. Given these bounds, these extensions of the SM are beginning to
suffer their own fine-tuning problem. In particular, the bounds on the
masses of the top quark partner make their crucial cancellation of UV
sensitivity less efficient. 

In most extensions of the SM mentioned above, the top partners carry
SM $SU(3)$ color.
However, it is possible to significantly alleviate the tuning of
extensions of the SM in models where the top partners are not charged
under the SM color gauge group. These theories, grouped under the name
of neutral natural models, mimic the SM extensions described
above. For the case when the Higgs is a pNGB, colorless top partners
are present in the Twin Higgs~\cite{Chacko:2005pe} and  Quirky Little Higgs~\cite{Cai:2008au} models. When
the symmetry protecting the Higgs mass is supersymmetry, this can be
realized in Folded Supersymmetry~\cite{Burdman:2006tz}. Many more
recent realizations of these ideas exist in the literature~\cite{Craig:2014aea,Craig:2014roa,Craig:2015pha,Barbieri:2015lqa,Low:2015nqa,Gherghetta:2016bcc,Craig:2016kue,Serra:2017poj,Cohen:2018mgv,Cheng:2018gvu}

In Folded Supersymmetry (FSUSY), just as in the other  neutral natural
models, there is an unbroken $SU(3)$  gauge group surviving at low
energies, folded color. The folded superpartners of the SM quarks, f-squarks, are
charged under this symmetry, not the SM color group. The minimal low energy
theory contains the SM and the f-squarks including the f-stop responsible
for stabilizing the   weak scale. In addition, since the lightest
matter states charged under f-color are very massive (with f-quarks in
the hundreds of GeV), the only light hadrons of the unbroken f-color
group are f-glueballs. These states, once produced cannot be detected,
with the notable exception of states that mix with the Higgs through a
top partner loop. Then once produced, the lowest lying $0^{++}$
f-glueballs will decay back to SM states with large lifetimes,
potentially resulting in highly displaced vertices.  The production of 
f-glueballs in rare decays of the Higgs have been studied in
Ref~\cite{Curtin:2015fna}.

 Here, we will be interested in the f-glueballs 
coming from  the annihilation of electroweakly produced f-squark pairs.
As shown in Ref.~\cite{Burdman:2008ek}, when f-squarks are produced
they cannot hadronize and must de-excite by shedding soft radiation
until they annihilate, mostly into f-gluon pairs. In turn, these will
hadronize into f-glueballs~\cite{Chacko:2015fbc}. Assuming that a
significant fraction of these f-glueballs will be in the $0^{++}$  ground 
state, they can decay back to the SM via higher dimensional operators. Here
we are interested in calculating the distribution of displaced
vertices that results from this process. For this purpose, we will need
to model the fragmentation of gluons into glueballs in the hidden
sector. Since the f-squark masses are much larger that the scale where
the f-color interaction becomes strong, the f-glueball mass
$M_{\tilde{G}}$ is expected to be considerably smaller that the
f-gluon energy. Then, the fragmentation process is important in
determining the pattern of displaced vertices  generated by each hard f-gluon. 

Although here applied to the specific case of FSUSY, the modeling of
the fragmentation function we perform generally applies to theories
with hidden glue  where matter is much heavier than the
hadronization scale. Thus, we expect that with only minor changes, the
fragmentation functions found here can be used to study the displaced
vertex signals of other hidden glue scenarios with glueballs such as
the fraternal twin Higgs ~\cite{Craig:2015pha}, the quirky little
Higgs~\cite{Cai:2008au}, the vector-like Twin
Higgs~\cite{Craig:2016kue}, the hyperbolic Higgs~\cite{Cohen:2018mgv},
a singlet-scalar top partner ~\cite{Cheng:2018gvu};  as well as any realization of hidden valley
models~\cite{Strassler:2006im,Strassler:2006ri,Han:2007ae} where
glueballs are the lightest hadron in the spectrum~\cite{Juknevich:2009ji,Juknevich:2010rhj}.
A better understanding of the fragmentation process into glueballs in
these well-motivated scenarios will allow us to test them at current experiments
and even in dedicated new proposed ones~\cite{Curtin:2018mvb}.

The rest of the paper is organized as follows: in the next section we
review general aspects of FSUSY and define the parameters of the
theory. In section~\ref{sec:pheno} we set up the phenomenology of 
f-squark production and its posterior annihilation into f-gluon pairs,
as well as the general aspects of the glueball lifetime. In
section~\ref{sec:dvs} we model the fragmentation function of f-gluons
into f-glueballs and obtain the distribution of displaced vertices  in
a variety of situations. Finally, we conclude in section~\ref{sec:conc}.

\section{Hidden Glue in Natural Extensions of the  Standard Model} 
\label{sec:glue}

In this section we will focus on FSUSY in order to provide a complete
example of the glueball parameters necessary to study the
phenomenology of the associated hidden glue. In particular, we need to
understand the relationship between the glueball mass and the matter
spectrum of the theory. In FSUSY, this is given by the f-squarks.
 
The aim of FSUSY is to solve the little hierarchy problem, i.e. the
already large gap between the weak scale and the scale of new physics
that is implied by the current bounds on new particles from the LHC as
well as indirect observables. In particular, since the top quark
contribution to $m_h^2$ is the largest, it determines the tuning of the
SM extension. In the typical weak scale supersymmetric models, this
means that the stop mass cannot be too large. The tension arises from
the direct bounds on squark and gluinos from the LHC, which are 
typically already above a TeV~\cite{Sopczak:2017ezw,Lacroixon:2018lcy}. The way FSUSY addresses this is by 
having a low energy spectrum where there are squarks that are not
charged under QCD, and not having gluinos. The relevant Yukawa terms
are 
\begin{equation}
{\cal L} \supset \left( Y_t h_u q_A t_A + {\rm h.c.}\right) +  Y_y^2
\left|\tilde{q}_B h_u\right|^2 + Y_t^2
\left|\tilde{t}_B\right|^2\,\left|h_u\right|^2~,
\label{eq:fyukawa}
\end{equation}
where the indices $A$ and $B$ refer to the SM QCD gauge group
$SU(3)_A$ and another $SU(3)_B$ gauge group under which the scalars
are charged. In (\ref{eq:fyukawa}) $q_A$ is the third generation quark
$SU(2)_L$ doublet and $h_U$ is the up-type Higgs doublet. It is clear
that this form of the Yukawa sector does not contribute quadratic UV
sensitivity to $m_h^2$, just as is the case of typical weak scale SUSY
models such as the minimal supersymmetric SM (MSSM). The crucial
difference is that the scalar top partners here do not carry QCD charges
and  therefore can only be electroweakly produced at colliders. It is
possible to obtain  (\ref{eq:fyukawa}) in UV completions in
five-dimensional (5D) theories where SUSY is broken by 
Scherk-Schwarz boundary conditions~\cite{Burdman:2006tz}.
The gauge symmetry is $SU(3)_A\times SU(3)_B\times Z_{AB}$, where
$Z_{AB}$ is a discrete symmetry that exchanges the $A$ and $B$
indexes. In the 5D construction the theory is supersymmetric in the
bulk. The Scherk-Schwarz SUSY breaking implements the orbifold
projection necessary to obtain the desired low energy spectrum: quarks
charged under $SU(2)_A$ and squarks charged under $SU(3)_B$, the
f-squarks. Besides the QCD-charged squarks, the orbifold procedure
gets rid also of gauginos, including gluinos. Then, as advertised, we
have the spectrum necessary to stabilize the little hierarchy without
introducing new states charged under the SM color.  

In addition to f-squarks, the minimal necessary FSUSY  low
energy spectrum includes  an additional unbroken $SU(3)_B$ gauge
interaction. Since the lightest matter fields charged under $SU(3)_B$
are the f-squarks, which are hundreds of GeV (see below), the low-lying
hadron spectrum of this new strong interaction is populated by
glueballs. 

The heavy matter charged under a new unbroken
non-abelian gauge interaction, in this case $SU(3)$, are generically
called quirks if fermionic or squirks if scalar. Their phenomenology
was studied in \cite{Kang:2008ea} and further in
\cite{Harnik:2011mv,Fok:2011yc}. 
The case of f-squarks was first studied in
Ref.~\cite{Burdman:2008ek}, where it is shown that after being produced
through electroweak interactions they will form a flux tube that emits
soft radiation, and eventually will come back to annihilate. All of
this process is prompt and results mostly into a pair of hard  f-gluons,
which will then hadronize into glueballs.  This will be the process of
interest for us in the rest of the paper. However, if we consider the
production through the charged current then the preferred annihilation
channel is into $W^\pm\gamma$. This leads to bounds on the left-handed
f-squark masses~\cite{Burdman:2014zta}  which for the LHC Run~I were
just below $500$~GeV.   The right-handed f-quarks are somewhat 
lighter since soft masses are generated by finite loop corrections
dominated by the gauge interactions~\cite{Delgado:1998qr}. We will then focus on
the glueballs generated by the annihilation of these f-quarks,
although much of what will do can also applied to the left-handed
squarks. 

\begin{figure}[tbp]
\centering % \begin{center}/\end{center} takes some additional vertical space
\includegraphics[width=0.8\textwidth]{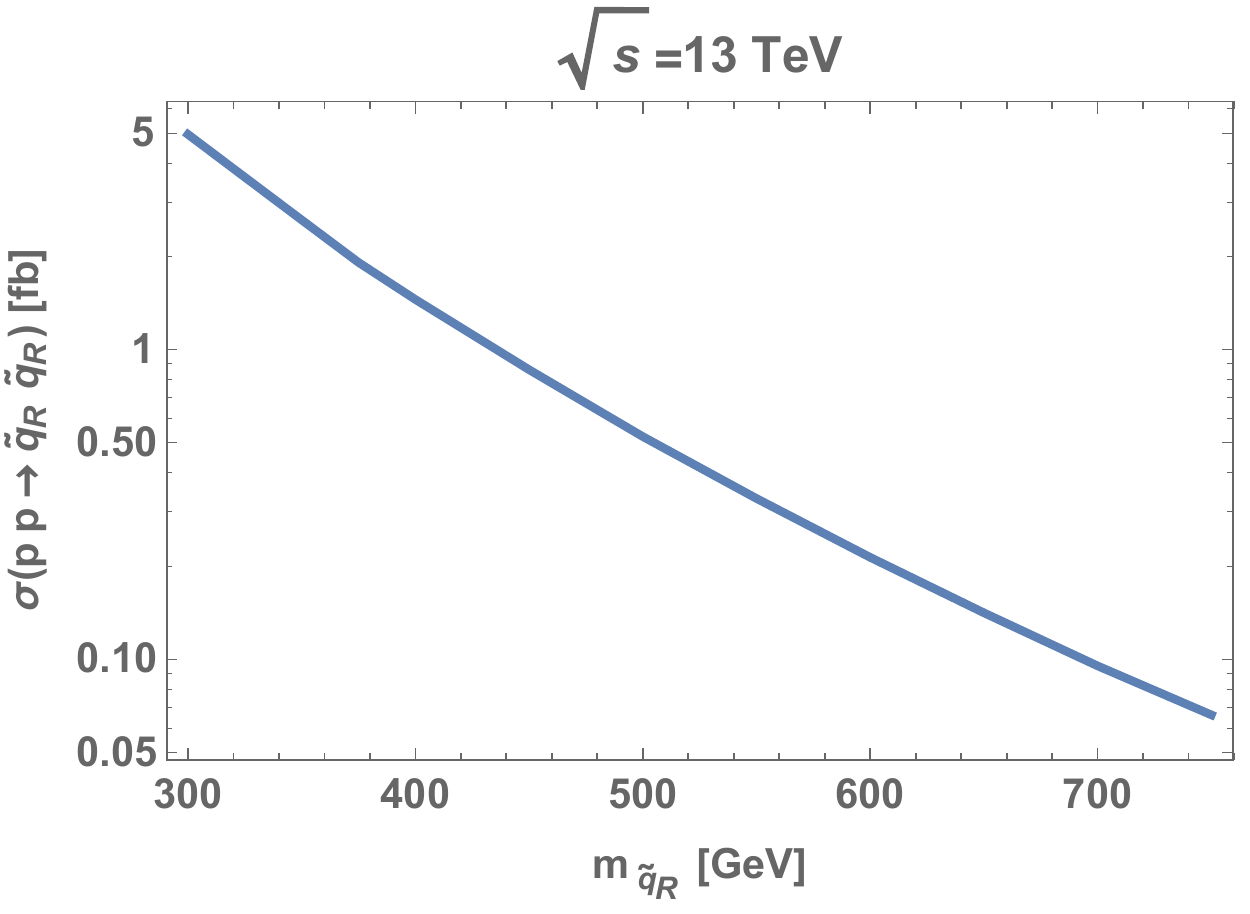}
% "\includegraphics" is very powerful; the graphicx package is already loaded
\caption{\label{fig:1} Cross section for the production of
  right-handed f-squarks vs. their mass. We assume
  $m_{\tilde{u}_R}=m_{\tilde{d}_R}=m_{\tilde{c}_R}=m_{\tilde{s}_R}=m_{\tilde{b}_R}$. See
  text for explanation.}
\end{figure}

The cross section for pair production of right-handed f-squarks at the
LHC with $\sqrt{s}=13$~TeV is
shown in Figure~\ref{fig:1} as a function of the f-squark mass. 
In the minimal F-SUSY realization the f-squark soft masses are
determined by finite loop effects~\cite{Delgado:1998qr}, mainly from
gauge interactions. Only the third generation doublet and the
right-handed top f-squarks are significantly affected by Higgs loops. Assuming this
holds in more general scenarios we take
\begin{equation}
m_{\tilde{u}_R}=m_{\tilde{d}_R}=m_{\tilde{c}_R}=m_{\tilde{s}_R}=m_{\tilde{b}_R}~,
\label{eq:degfsqmass}
\end{equation}
only excluding the right-handed f-stop. Furthermore, the left-handed
f-squarks are heavier due to the additional contributions coming form
the $SU(2)_L$ sector~\cite{Burdman:2006tz}. So here we consider the
production of the lighter degenerate right-handed f-squark sector
excluding the f-stop. This will give us a baseline to obtain estimates
of glueball production. 

Once pair-produced, the f-squarks  fly apart but they cannot hadronize
due to the absence of light f-colored matter. The ensuing process of
de-excitation has been studied in detail in
Ref.~\cite{Burdman:2008ek}. After shedding some soft electromagnetic and 
f-glueball radiation, the pair forms a s-wave f-squarkonium, which
then decays either to f-gluons or to a lesser extent back to SM
fermions or gauge bosons. 
The charged pair production through an s-channel $W^\pm$ was studied
in ~\cite{Burdman:2014zta}. In this case the left-handed f-quarks
decay back preferentially to $W^\pm\gamma$, resulting in bounds on
$m_{\tilde{q}_L}$. The current bounds vary from $320$~GeV to $465$~GeV
extracted from the $8$~TeV ATLAS data~\cite{Aad:2014fha}, where the range
corresponds to considering one to three generations of left-handed
f-squarks. 

Right-handed f-squarks are expected to be lighter. However,
since there main decay mode is into glueballs, at  the moment we lack any bounds on
their masses.  In these cases the annihilation of f-squarks after they
are pair-produced and they come back together, is expected to
overwhelmingly go into a pair of f-gluons, with  each of them
hadronizing to a number of 
glueballs. Below we will model this hadronization process. Once
produced, folded glueballs would escape detection unless they decay
back to SM states. This has been shown to happen through Higher
dimensional operators induced by the Higgs first in
Ref.~\cite{Craig:2015pha} in the context of the Fraternal Twin Higgs
(FTH) model, and later applied also to FSUSY in
Ref.~\cite{Curtin:2015fna,Chacko:2015fbc}. This will allow us to
compute the glueball lifetime, which together with the fragmentation
function will result in a distribution of displaced vertices. The main
parameter needed to compute  both the fragmentation function  and the
decay of glueballs is their mass.  This is largely determined by the
infrared scale defined by the hidden glue interactions, $\Lambda_{\rm
  IR}$. In turn, this requires knowledge of the spectrum of states
charged under the hidden glue up
to a $UV$ scale. In the case of FSUSY, this means knowing the f-squark
spectrum up to the scale where supersymmetry is restored, which we
call $\Lambda_{\rm UV}$. This will give us the running of the hidden
glue $SU(3)_B$ coupling, $\alpha_F(q^2)$, with the $UV$ boundary
condition  
\begin{equation}
\alpha_F(\Lambda_{\rm UV}) = \alpha_s(\Lambda_{\rm UV})~,
\label{uvbc}
\end{equation}   
which is a $Z_2$-preserving UV boundary condition. 
In \cite{Burdman:2006tz} the FSUSY low energy spectrum was
UV-completed by a flat compact extra dimension, which was used to
break supersymmetry at the compactification scale $1/R$ by the
Scherk-Schwarz boundary conditions.  The f-squark soft masses then are
generated by gauge loops of size $1/R$~\cite{Delgado:1998qr}, as
mentioned above. This gives a minimal way to generate the soft masses
and in principle there could be additional (localized) supersymmetry breaking
sources changing these soft masses. However we will assume that these
would not break the relation between left- and right-handed squark
masses, which is approximately given by
\begin{equation}
m^2_{\tilde{q}_L} \simeq  m^2_{\tilde{q}_R}
+\frac{k}{4\pi^4}\,\frac{3}{4}\,g^2\,\Lambda^2_{\rm UV}~,
\end{equation}
where we are using $\Lambda_{\rm UV} = 1/R$, we are neglecting terms
of order $g'^2$, and $ k\simeq 2.1$ is computed in the extra dimension
UV-completion~\cite{Delgado:1998qr}. Then, even if we do not make
detailed assumptions about supersymmetry breaking, we can compute
$\alpha_F(q^2)$ for a given $\Lambda_{\rm UV}$ and $m_{\tilde{q}_R}$. 
We use this relation to compute the evolution of $\alpha_F(q^2)$ with
the boundary condition (\ref{uvbc}) in order to obtain $\Lambda_{\rm
  IR}$. Finally, we estimate the lowest lying $0^{++}$ glueball mass making  use of the approximate relation 
\begin{equation}
M_{\tilde{G}} \simeq 7\,\Lambda_{\rm IR}~,
\label{glueballmass}
\end{equation}
which has been observed to hold in
QCD~\cite{Morningstar:1999rf,Chen:2005mg,Gockeler:2005rv,Craig:2015pha}. The
$0^{++}$ state can decay back to SM states through its mixing with the
Higgs boson. Heavier states, such as the $2^{++}$, the $0^{-+}$, etc.,
will also be produced in the folded gluon fragmentation process, but
they would escape detection. We will then focus here on the
fragmentation and decay properties of the $0^{++}$ glueballs. 

\begin{figure}[tbp]
\centering % \begin{center}/\end{center} takes some additional vertical space
\includegraphics[width=0.8\textwidth]{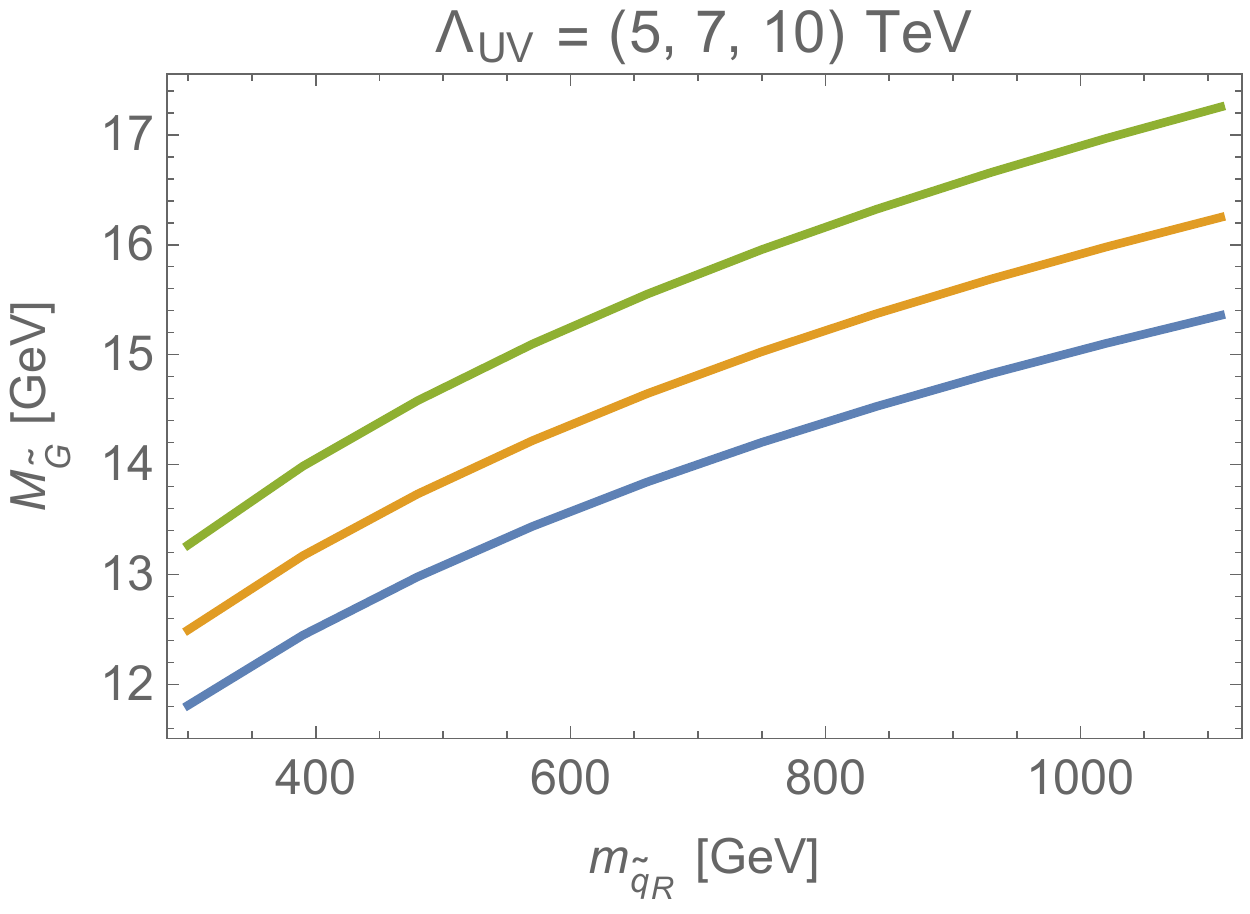}
% "\includegraphics" is very powerful; the graphicx package is already loaded
\caption{\label{fig:2} $0^{++}$ glueball mass as a function of the
  right-handed f-squark mass, for (from bottom to top)  $\Lambda_{\rm UV}=5, ~7,~10~$TeV. 
}
\end{figure}

Assuming the spectrum is approximately degenerate across generations
we can then compute $M_{\tilde{G}}$ as a function of
$m_{\tilde{q}_R}$. For instance in Figure~\ref{fig:2} we show the
  $0^{++}$ glueball mass for typical values of the UV cutoff
  $\Lambda_{\rm UV}$. 
We see that folded  glueball masses  are typically
$(12-20)~$GeV range for the interesting values of
$m_{\tilde{q}_R}$. This gives us an estimate of the relation between
$M_{\tilde{G}}$ and the typical folded gluon energy fragmenting into
glueballs, which we will use in the next section to model the
fragmentation function.

\section{Glueball Fragmentation } 
\label{sec:pheno}
We are interested in modeling the fragmentation of energetic
f-gluons (resulting from either f-squark annihilation or even
from rare Higgs decays) into folded glueballs. 
In this section we lay the groundwork for this in a generic way that
should mostly applied to other examples with hidden glue sectors
resulting in glueballs.  
In general, the fragmentation function $D_i^H(E_h/E_i)$ of a given parton  $i$ into a
hadron $h$ obeys 
\begin{equation}
\frac{d\sigma(pp\to h+X)}{dE_h} = \sum_i \frac{d\sigma(pp\to i
  +X)}{dE_i}\, D_i^h(E_h/E_i)\,\frac{dE_i}{E_i}~,
\label{fragfundef}
\end{equation}
where $E_i$ and $E_h$ are the parton and  hadron energies respectively,
and we sum over all possible partons. Here we will consider only one
parton, the folded gluon $\tilde{g}$, as well as only one hadron, the
$0^{++}$ glueball, or $\tilde{G}$. Fragmentation to higher mass states
such as $2^{++}$ and $0^{-+}$ that do not mix with the Higgs would result in missing energy since
they escape detection. For our purposes, we only need to assume that
the fragmentation fraction into the $0^{++}$ state is constant with
energy. This is a good approximation for f-gluon energies much greater
than $M_{\tilde G}$, which is the case here. Since here we are mainly
interested in the form of the distribution of displaced vertices this
will not affect our results (see below) , which just  need only be multiplied by
this constant, $r_{\tilde G^0}$. Estimates~\cite{Juknevich:2010rhj} using a thermal model for a
pure glue theory give $r_{\tilde G^0}\simeq (0.5-0.6)$. On the other
hand, the results can be affected by the presence of soft f-gluon
radiation in between the two hard f-gluons. These would result in a
few  additional, slower moving f-glueballs. Since our calculation only
considers the fragmentation from the hard f-gluons, it should be
considered a lower limit to the number of displaced vertices for a
given value of $r_{\tilde G^0}$. 

 Defining the glueball momentum
fraction $x= E_{\tilde{G}}/E_{\tilde{g}}$,  the availability  of only one
hadron for the gluon to fragment into results in the simple
energy-conservation constraint
\begin{equation}
\int_{x_{min}}^1 dx\, x\, D_{\tilde{g}}^{\tilde{G}}(x) = 1~,
\label{sumrule1}
\end{equation}
where 
\begin{equation}
x_{min} = \frac{M_{\tilde{G}}}{E_{\tilde{g}}}~.
\end{equation}
In general, fragmentation functions are parametrized and fit to data at
some energy, and then they are  evolved using the DGLAP evolution
equations in order to use them at different energies.  This method is
not available to us since we are trying to predict the glueball
signal. However we can model 
$D_{\tilde{g}}^{\tilde{G}}(x)$ at a given input energy $s_0$ with a simple parametrization and impose
  constraints to reduce our ignorance of the fragmentation process.
  We start with the simple two-parameter form
\begin{equation}
D_{\tilde{g}}^{\tilde{G}}(x) = N_v\,\left(1-x\right)^\beta~,
\label{largexparam}
\end{equation}
where $N_v$ and $\beta$ are free parameters, and we should read this as $D_{\tilde{g}}^{\tilde{G}}(x,s_0)$. This simple expression
encodes the fact that we expect the fragmentation function to fall to
zero for values of $x$ not too far from $1$. 

 However, this is not a
good parametrization at low $x$. For one, the fragmentation should
also fall to zero at very low $x$, where it is expected that is
dominated by singularities in the gluon splitting function. 
The DGLAP evolution of  $D_{\tilde{g}}^{\tilde{G}}(x,s)$ is given by 
\begin{equation}
s\,\frac{\partial D(x,s)}{\partial s} = \int_x^1
\frac{dz}{z}\,\frac{\alpha_F(s)}{2\pi}\,P_{\tilde{g}\tilde{g}}(z,\alpha_F(s))\,D(x/z,s)~,
\label{dglap}
\end{equation}
where for simplicity we denoted the f-gluon to f-glueball
fragmentation function as $D(x,s)$, and the evolution will only depend
on the folded gluon splitting function
$P_{\tilde{g}\tilde{g}}$. If we use the  corresponding gluon-gluon QCD
splitting function to NLO~\cite{Furmanski:1980cm}, the 
singularities in $P_{\tilde{g}\tilde{g}}$ at low $x$ in (\ref{dglap})
dominate and result in~\cite{Ellis:1991qj} 
\begin{equation}
D(x,s) \propto \frac{1}{x}\,e^{-\frac{(\xi-\xi_p)^2}{2\sigma^2}}~,
\label{lowxd}
\end{equation} 
where
\begin{equation}
\xi\equiv \ln\frac{1}{x}\qquad \xi_p\equiv
\frac{\pi}{b\alpha_F(s)}~\qquad \sigma\equiv \left(\frac{\pi}{6b}\sqrt{\frac{2\pi}{C_A\,\alpha_F^3(s)}}\right)^{1/2}~,
\label{xidefs}
\end{equation}
with $b$ the coefficient of the folded $SU(3)_B$ color 
beta-function. The low $x$ form of the fragmentation function is valid up
to about 
\begin{equation}
\ln\frac{1}{x} \simeq \frac{\pi}{b\alpha_F(s)}~.
\label{xmatch}
\end{equation}
Thus, we will use this form of $D(x,s)$ up to some appropriate value
of $x$ and then match it to the form in (\ref{largexparam}) valid for
larger values of $x$. 

An additional constraint on the fragmentation function is that its
integral results in the average glueball multiplicity $\langle n\rangle$. This is 
\begin{equation}
\int_{x_{min}}^1 dx\,D(x,s) = \langle n(s)\rangle~.
\label{avgmulti}
\end{equation}
Then, a strategy to model the fragmentation of gluons into glueballs
can be the following: start at an arbitrary energy scale $s_0$. Then  choose
a value of the power $\beta$ in (\ref{largexparam}). This choice will
turn out to be equivalent to measuring the average multiplicity at
$s_0$. Then, we match the two forms (\ref{largexparam}) and (\ref{lowxd})  at some appropriate value of $x$
and  finally impose energy conservation (\ref{sumrule1}). This
procedure determines $D(x,s_0)$ (and also $\langle
n(s_0)\rangle$). Then we can use the DGLAP evolution (\ref{dglap}) to
obtain $D(x,s)$ at any other desired energy scale.

\begin{figure}[tbp]
\centering % \begin{center}/\end{center} takes some additional vertical space
\includegraphics[width=0.8\textwidth]{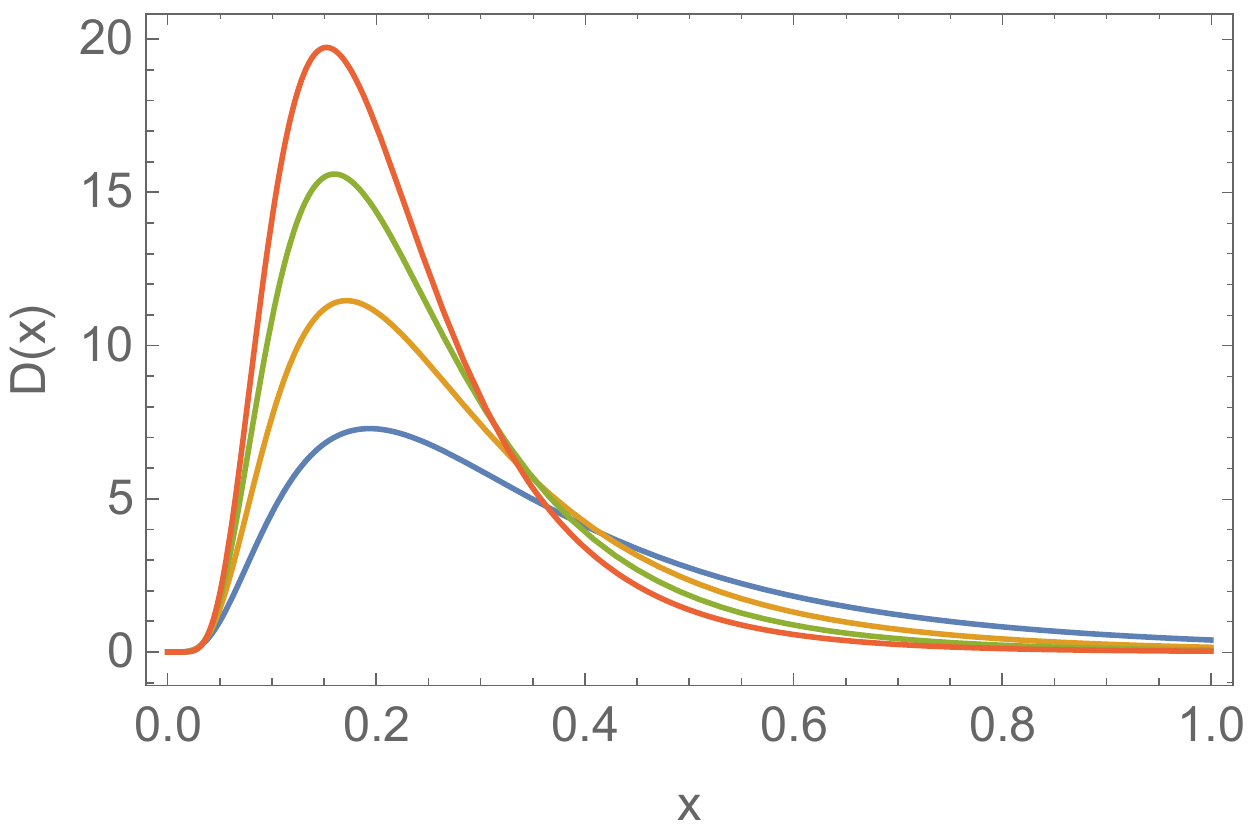}
% "\includegraphics" is very powerful; the graphicx package is already loaded
\caption{\label{fig:3} 
Fragmentation function as a function of the energy fraction $x$ for
various energies. From bottom to top: $300~$GeV, $500~$GeV, $700~$GeV
and $900~$GeV. Here we fix the matching value as $x_M=0.1$, and used
$M_{\tilde G}=15~$GeV. 
}
\end{figure}

It will be convenient  to impose the DGLAP evolution directly on the
average multiplicity. The
key observation is that the average multiplicity is completely
dominated by small values of $x$. As a result it is an excellent
approximation to use the following form of the energy dependence for
the average multiplicity~\cite{Ellis:1991qj}
\begin{equation}
\langle n(s)\rangle \propto
e^{\frac{4\pi}{b}\sqrt{\frac{6\pi}{\alpha_F(s)}}+\frac{1}{4}\ln\alpha_F(s)}~,
\label{nofs}
\end{equation} 
where we have evaluated the exponential for zero active flavors,
$n_f=0$. Imposing this energy dependence, and starting from a fixed
value $\langle n(s_0)\rangle$, we fit the low and high $x$
parameterizations (\ref{lowxd}) and (\ref{largexparam}) to obtain $D(x,s)$.
Imposing a value of the average multiplicity at $s_0$ is equivalent to
choosing an initial value of $\beta$ in
(\ref{largexparam}). The result of this process starting with
$\beta=1$ at $s_0=200~$GeV is shown in Figure~\ref{fig:3}.
 We see that as the reference energy increases, the large $x$
 contributions are suppressed in favor of the low $x$. In the language
 of the parametrization (\ref{largexparam}) this corresponds to
 increasing the power $\beta$.

\begin{figure}[tbp]
\centering % \begin{center}/\end{center} takes some additional vertical space
\includegraphics[width=0.8\textwidth]{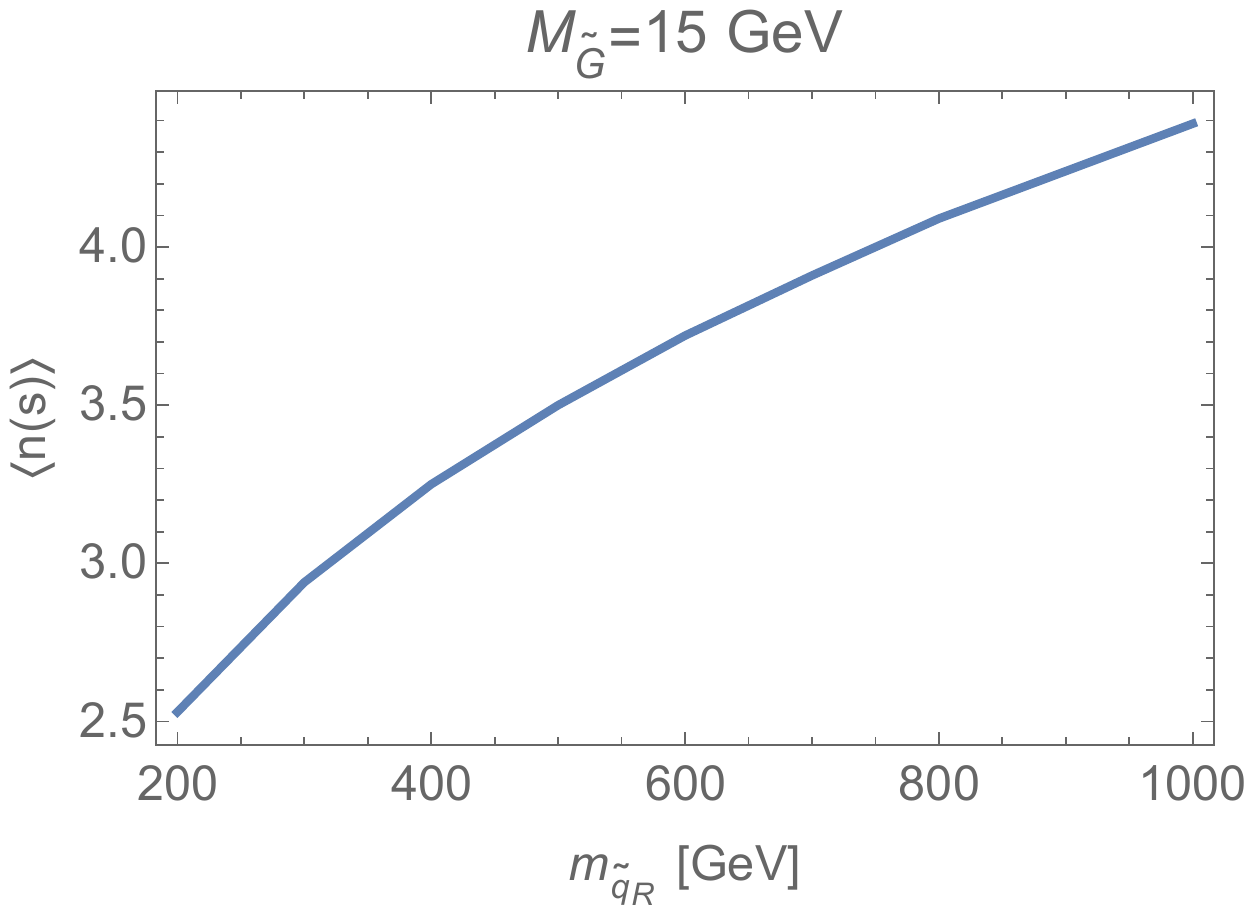}
% "\includegraphics" is very powerful; the graphicx package is already loaded
\caption{\label{fig:4} 
Average multiplicity of glueballs as a function of energy. Here we used
$x_M=0.1$ and $M_{\tilde G}=15~$GeV. 
}
\end{figure}

The corresponding average multiplicity as a function of the energy is
shown in Figure~\ref{fig:4}. 
The input value is taken to be $\langle n(s_0)\rangle\simeq 2.5$ for 
$s_0=200~$GeV, which corresponds to choosing $\beta=1$ for large $x$
at that energy. We choose $s_0$ to have a value safely higher than any
possible value for $M_{\tilde{G}}$.  Although this value for the
parameter $\beta$ at $s_0$, and
therefore  the normalization of the average multiplicity 
is a priori arbitrary, it reflects the simplest parametrization of the
large momentum fraction region, i.e. a linearly falling
fragmentation. 

The choice of $\beta\simeq 1$ for $s_0$ is also suggested by the similarity observed between the
glueball wave-function and that of the $\pi^0$, as obtained in QCD sum
rules~\cite{Wakely:1991eu,Wakely:1991ej}.  This implies that the
fragmentation function of a gluon into a glueball is similar to the
one of valence quarks into $\pi^0$, as it is pointed out  in~\cite{Roy:1999bu}.
The relevant fits to the parametrization (\ref{largexparam}) valid at
large $x$ can be found, for instance, in ~\cite{Chiappetta:1992uh},
where we see that $\beta\simeq 1$ correctly reproduces the average multiplicity.
The more sophisticated parameterizations~\cite{Patrignani:2016xqp}  currently used, for instance,
for the pion fragmentation function are next-to-leading order and are
designed to be valid for both small and high momentum fractions. 
Ultimately, this input parameter  can be fixed by using a Monte Carlo simulation of
the parton shower in a quarkless model of QCD, i.e. a simulation of
the F-SUSY low energy theory. This would greatly reduce the
uncertainty in the model for $D(x)$. In the meantime, having a model
for the fragmentation allows us to make  estimates of the qualitative
behavior of the signal in detectors.  

\begin{figure}[tbp]
\centering % \begin{center}/\end{center} takes some additional vertical space
\includegraphics[width=0.8\textwidth]{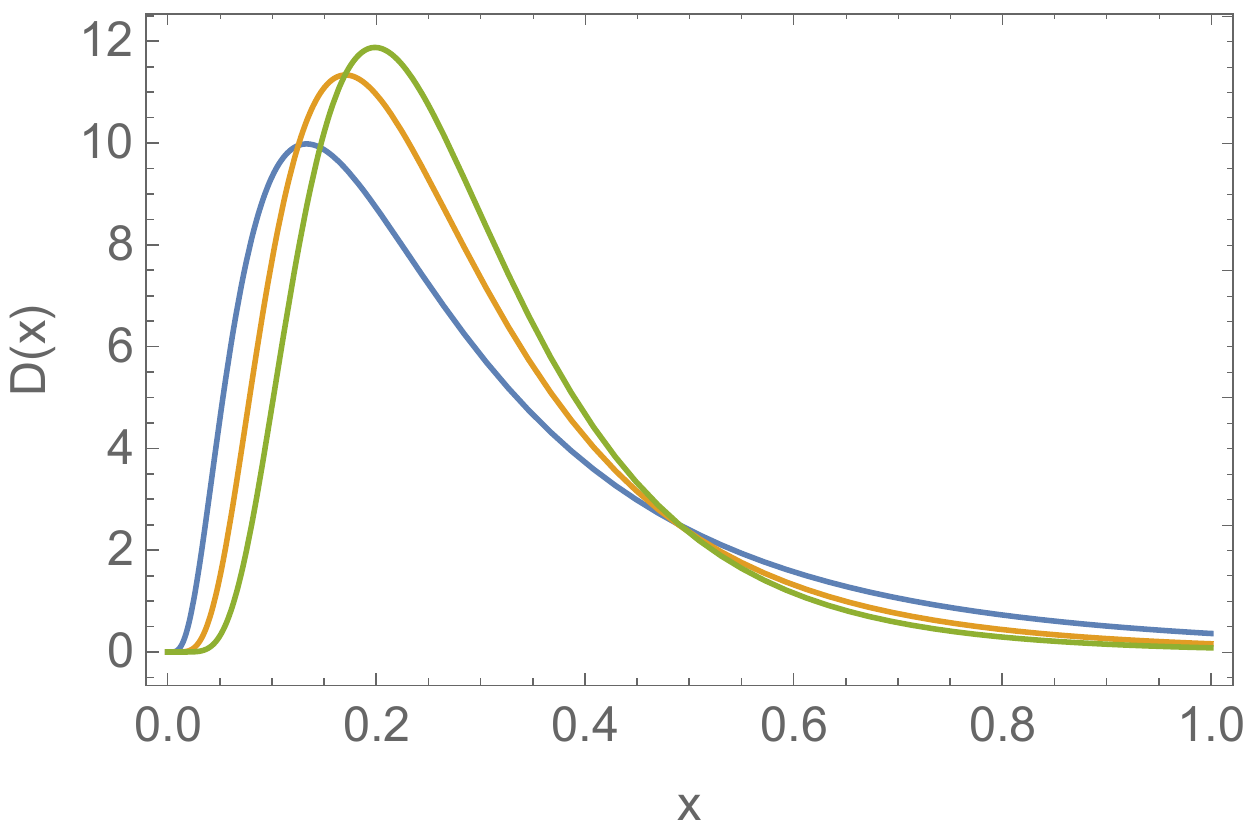}
\caption{\label{fig:5} 
Fragmentation function as a function of the energy fraction $x$ for
$500~$GeV, for various choices of matching value $x_M$.
 From bottom to top: $x_M=0.07~$, $0.10~$ and  $0.13~$.
}
\end{figure}

One remaining arbitrary parameter of the model is the value of the
momentum fraction  $x_M$ used to match the low and high $x$ behaviors
of the fragmentation function. In Figure~\ref{fig:5} we vary its value
in order to study the sensitivity of the result to this choice. We
fix the gluon energy at $500~$GeV and use for the matching momentum
fraction three values $x_M=0.07, 0.10, 0.13$. From the figure we see
that the systematic uncertainty associated with this parameter is not
too large, especially if we consider this as a useful model of the
fragmentation function that will allow us to  simulate the distribution
of the displaced vertices in a detector. To quantify the uncertainty
we look at the average multiplicity resulting from these three
choices: $\langle n(500~{\rm GeV})\rangle = 3.60, 3.50, 3.36$, and see
that it is not such a large variation.
As we will see in the next section, the largest uncertainty in the
prediction for the distribution of displaced vertices from glueball
production and decays will be the glueball mass.

\section{Displaced Vertices from Glueballs}
\label{sec:dvs}
\begin{figure}[tbp]
\centering % \begin{center}/\end{center} takes some additional vertical space
\includegraphics[width=0.6\textwidth]{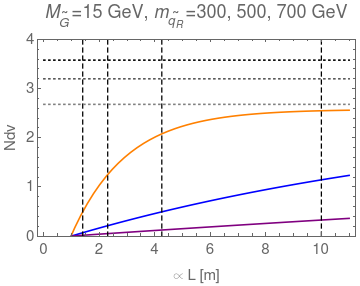}
\caption{\label{fig:6} 
Distribution of displaced vertices for $m_{\tilde q}=300,500$ and
$700~$GeV, and $M_{\tilde G}=15 ~$GeV. The vertical axis, Ndv is the
number of displaced vertices as a function of distance L from the
primary, normalized by $r_{\tilde G^0}$, the f-gluon fragmentation
fraction into the $0^{++}$ glueball.
The vertical lines indicate
approximate boundaries of the inner tracker, electromagnetic
calorimeter, hadronic calorimeter and muon system of the ATLAS
detector. The horizontal dotted lines correspond to the average
multiplicity for each of the three cases for the squark masses, from
 bottom to top: $300~$GeV, $500~$GeV and $700~$GeV.
}
\end{figure}

\begin{figure}[tbp]
%\centering % \begin{center}/\end{center} takes some additional vertical space
\begin{center}
\includegraphics[width=0.45\textwidth]{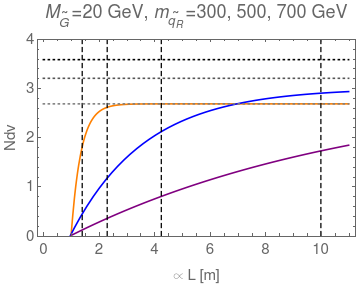}\quad
\includegraphics[width=0.45\textwidth]{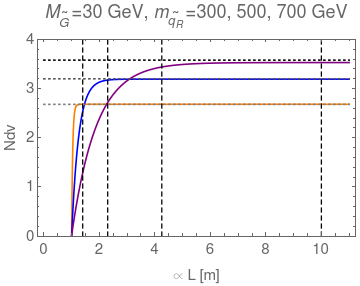}
\caption{\label{fig:7} Same as Figure~\ref{fig:6}, but for 
 $M_{\tilde G}=20 ~$GeV (left) and  $M_{\tilde G}=30 ~$GeV (right). 
}
\end{center}
\end{figure}

Armed with the gluon-to-glueball fragmentation function $D(x,s)$ we
are now in the position to predict the distribution of displaced
vertices generated once they decay back to SM states.  
Assuming that the gluon fragments into a glueball promptly, the number
of displaced vertices as a function of the decay length $L$ is given by
\begin{equation}
N_{\rm dv} (L) = \int_{x_{\rm min}}^1 dx\,D(x,s) \,
\left(1-e^{L/\bar{L}_{\tilde G}(x)}\right)~,
\label{ndv}
\end{equation}
where we have defined 
\begin{equation}
\bar{L}_{\tilde G}(x)\equiv c\,\tau_{\tilde G}\,\frac{x}{x_{\rm min}}~,
\label{lbar}
\end{equation}
which is the decay length determined by the glueball lifetime
$\tau_{\tilde{G}}$. As seen in (\ref{lbar}), it depends on the energy
fraction of the glueball, $x/x_{\rm min} $ which is simply the $\gamma$
factor. In order to compute the distribution in (\ref{ndv}) we need to
know the glueball lifetime, which depends crucially on details of the
model since $\tau_{\tilde{G}}$ is typically determined by a
higher dimensional operator mixing the $0^{++}$ glueball with the
Higgs boson. This is given by~\cite{Juknevich:2009ji,Craig:2015pha}
\begin{equation}
{\cal O} \sim \frac{\alpha_F}{4\pi}\,
|H|^2\,{\rm Tr}\left[\tilde{G}_{\mu\nu}\tilde{G}^{\mu\nu}\right]~,
\label{dim6op}
\end{equation}
with $\tilde{G}_{\mu\nu}$ the hidden gluon field strength.
For the case of FSUSY the coefficient multiplying (\ref{dim6op}) is given by \cite{Curtin:2015fna}
\begin{equation}
\frac{1}{48\pi}\,\frac{y^2_t}{m^2_{\tilde{t}}}~,
\label{dim6opcoeff}
\end{equation}
where $y_t$ is the top
Yukawa coupling and  it is assumed that
$m_{\tilde{t}_1}=m_{\tilde{t}_2}=m_{\tilde{t}}$. Then, the resulting glueball decay width to a given SM final
state  is 
\begin{equation}
\Gamma(\tilde{G}\to {\rm SM}) =
\left(\frac{\alpha_F}{48\pi}\,\frac{y_t^2}{m^2_{\tilde{t}}}\,\frac{v}{m_h^2-M^2_{\tilde{G}}}\right)^2
\,f^2_{\tilde{G}} \,\Gamma(h\to {\rm SM})~.
\label{gbtosm}
\end{equation}
Here, the Higgs decay width to the SM final state corresponds to that
of an off-shell Higgs, and it must be evaluated with the replacement
$m_h\to M_{\tilde{G}}$. Also in (\ref{gbtosm}) we defined the hadronic
matrix element creating the glueball from the gluon operator defined
by 
\begin{equation}
\langle 0| {\rm
  Tr}\left[\tilde{G}_{\mu\nu}\tilde{G}^{\mu\nu}\right]|\tilde{G}\rangle
  \equiv f_{\tilde{G}}~,
 \label{gbdecayconst}
\end{equation}
with $f_{\tilde{G}}$ the glueball decay constant. Lattice studies
result in~\cite{Meyer:2008tr} 
\begin{equation}
4\pi\alpha_F \, f_{\tilde{G}} \simeq 2.3\,M^3_{\tilde{G}}~.
\label{latticefgb}
\end{equation}
Since the Higgs decay width back to the SM goes like $\sim
M_{\tilde{G}}$ we see that the lifetime scales as 
\begin{equation}
\tau_{\tilde{G}} \sim \frac{m^4_{\tilde{t}}}{M^7_{\tilde{G}}}~.
\label{tauscaling}
\end{equation}  
Thus, the glueball lifetime is extremely sensitive to the details
resulting in the glueball mass. To illustrate this point we show the
distribution of displaced vertices as a function of the distance to
the glueball production point for three values of the glueball mass:
$M_{\tilde G}= 15, 20$ and $30~$GeV in Figures~\ref{fig:6}
and ~\ref{fig:7}. In Figure~\ref{fig:6} we plot the number of
displaced vertices as a function of the distance from the production
point for $M_{\tilde G}=15~$GeV, for three values of the masses of the
squarks whose annihilation gives rise to the hard folded gluons
hadronizing into glueballs. The vertical lines correspond to the
approximate positions of the different sections of an idealized  version
of the ATLAS detector\footnote{This is just chosen for
  illustration. We do not attempt here a realistic detector
  simulation. An analogous study can be done for an idealized version
  of CMS.} 
starting from the trackers, then the electromagnetic and hadronic
calorimeters and all the way to the muon systems going out to about
$10$ meters. The horizontal lines are the average multiplicities for
each value of the squark mass. For instance, we see that even for
relatively small masses such as  $m_{{\tilde q}_R}=300~$GeV an
important fraction of the displaced vertices occur in or beyond the
muon system (about 1/3), with about half of the events in the
calorimeters. For the larger squark masses, most of the events within
the detector will be in the muon system, although the majority
of the displaced vertices will be beyond it.      

The situation changes drastically as the glueball mass is
increased. In Figure~\ref{fig:7} we show the distribution of displaced
vertices for $M_{\tilde G}=20~$GeV (left)  and  $M_{\tilde G}=30~$GeV
(right). For instance, for   $M_{\tilde G}=20~$GeV we see that for
$m_{{\tilde q}_R}=300~$GeV almost all displaced vertices would be in
the electromagnetic calorimeter. For the most massive case,
$m_{{\tilde q}_R}=700~$GeV, almost 2/3 of the events are inside the
detector, with roughly half of them in the calorimeters and the other half in
the muon system. For $M_{\tilde G}=30~$GeV, we see that all of the
events are inside the detector even for the larger squark masses considered,
with a large fraction in the electromagnetic calorimeter. 
Thus, even relatively small changes in the glueball mass would lead to a  very
different distribution of displaced vertices. This calls for
developing strategies for searching for displaced vertices in all
segments of the LHC detectors. In particular, since folded glueballs in FSUSY 
decay preferentially to a pair of b quarks, detections of these
should be made possible in all detector components, not just the
hadronic calorimeter. 

There can be other theories with glueballs or other low lying states with long
lifetimes induced, for instance, by higher dimensional operators such
as in FSUSY. Although  some of the details may be different, the
procedure to obtain the fragmentation function and then the
distribution of displaced vertices is analogous to what was done
above. 

Depending on the folded gluon energy resulting from the f-squark
annihilation, as well as on the glueball mass, the number of displaced
vertices inside the detector {\em on each side} of the event can be
larger than one. This presents a challenge for the searches for these
events at ATLAS and CMS. The modeling of the glueball fragmentation
function allows us to have a more detailed knowledge of the multiplicity
and of the location of the displaced vertices as a function of the
parameters of the models.

\vskip0.5cm

Searches for  long-lived particles need to include the
possibility of having the kind of secondary vertex topology that
appears in these models. 
For instance in \cite{Aad:2015uaa} the ATLAS collaboration considers
three kinds of models: a singlet scalar decaying to two long-lived
scalars each decaying to two jets, stealth SUSY and a hidden-valley
model where a $Z'$ decays to several long-lived scalar. In the first
two cases the signal contains two displaced vertices with two-jet
pairs which are somewhat back to back. This search requires two
displaced vertices (either in the inner detector or in the muon system
or one in each). This situation is reproduced in the folded gluon
fragmentation into glueballs for rather light f-squarks and small
glueball masses (e.g. top curve in Fig~\ref{fig:6}, for
$m_{\tilde{q}_R}=300~$GeV), although the decays tend to happen in the
calorimeters instead. On the other hand, for larger values of
$m_{\tilde{q}_R}$ (bottom two curves of Figure~\ref{fig:6}), there
maybe even less than two displaced vertices inside the detector. In these cases 
it is important to look for large values of missing transverse
energy. 

When there are more than two displaced vertices in
the event such as in the second half of the top curve of
Figure~\ref{fig:6} or in Figure~\ref{fig:7}, the  high
multiplicity  may result in failing isolation cuts. Although in these
cases it  is possible that some of the decays are outside the detector
and give large missing transverse energy, it may be also possible to
make use of the topology of the event which is still dominated by two
well separated clusters almost back to back resulting from the very
energetic folded gluons. A combination of the reconstruction of some
of the displaced vertices, the topology of the multi-jet event and
some missing $E_T$ may be needed to isolate these events from
background. In any case, an increased efficiency for the
reconstruction of displaced vertices in all detector segments might be
necessary, particularly in the muon system. 

On the other hand, the CMS collaboration~\cite{Sirunyan:2017jdo}
has done a study using a benchmark model with the pair production of
two long-lived scalars decaying to jets similar to the first model
considered by ATLAS in \cite{Aad:2015uaa}. However, in this analysis
displaced vertices are not fully reconstructed but identified by
``displaced jets'' variables. These are used as selection criteria for
jets, and then each event selected will contain a number of jets
tagged as displaced. This study is, in principle, not limited by the
number of tagged jets to a fix number of displaced vertices and could
be an adequate framework for the case with many folded glueballs.
More detailed studies following the experimental strategies and making
use of the information from fragmentation are left for future work.

\section{Conclusions}
\label{sec:conc}
Theories with hidden strong sectors may result in low lying states
without interactions with SM matter, but which can decay back to the SM via
higher dimensional operators. In this paper we studied the case of
FSUSY glueballs produced through the annihilation of electro-weakly
produced folded squark pairs. The f-squarks annihilate into hard 
f-gluons which then hadronize into f-glueballs. 
Although in general f-gluons would hadronize into several f-glueball
states we have only considered the lowest lying one, $0^{++}$, since
it can be detected via its mixing with the Higgs.  The fragmentation
fraction into $0^{++}$, $r_{\tilde G^0}$, is expected to be a constant
with energy for the case we study here. Our results for the number of
displaced vertices should then be multiplied by it, which
can be estimated using a thermal model to be $r_{\tilde G^0}\simeq (0.5-0.6)$.
 
We have modeled the glueball
fragmentation process obtaining a fragmentation function.   This was
achieved by appropriately parameterizing the low and high momentum fraction regions
of the fragmentation function. At low momentum fractions the
parametrization is chosen to be compatible with the known behavior of
the average multiplicity in (\ref{nofs}).  A simple parametrization
for the region of high momentum fraction, that of (\ref{largexparam}), is
used as initial input. Imposing energy conservation through (\ref{sumrule1})
and the average multiplicity (\ref{avgmulti}) we fix all parameters at a
given input energy $s_0$ except
one. We take this to be the value of $\beta$ in (\ref{largexparam}) at
the input energy $s_0$, which determines $\langle n(s_0)\rangle$ by
fixing the overall energy-independent constant in (\ref{nofs}). Once
this is done, the energy dependence of the fragmentation function
$D(x,s)$  is fixed by
DGLAP evolution. This is taken into account by imposing (\ref{nofs})
on the integral of the fragmentation function $D(x,s)$ resulting in
the average multiplicity, equation (\ref{avgmulti}). This translates
into a DGLAP evolution of the parameters for $D(x,s)$ in both the
small as well as the large momentum fraction regimes. The scale where
these  two parameterizations are matched, $x_M$ is also free parameter
but is approximately given by (\ref{xmatch}), which determines the
validity of the low momentum-fraction region. As seen in
Figure~\ref{fig:5}, variation of $x_M$ about a reasonable range
compatible with (\ref{xmatch}) does not result in dramatic changes in
the fragmentation function. Thus, the main parametric freedom can be
traced back to the overall normalization of (\ref{nofs}), or the
reference value of the parameter $\beta$ in (\ref{largexparam}) for a
given energy $s_0$. We have used $\beta=1$ for $s_0=200~$GeV, which
results in reasonable values of the average multiplicity. This linear
parametrization for the  reference value is inspired by a similar one
successfully used for the neutral  pion. Ultimately, this
arbitrariness can be removed by using a parton shower simulation in
FSUSY to compute the average multiplicity at the reference energy
$s_0$. This would result in a much less uncertain glueball
fragmentation function, which can then be used with more confidence to
extract model information from events with displaced vertices. 

In Section~\ref{sec:dvs} we have made use of these folded glueball
fragmentation functions to obtain the distribution of displaced
vertices at a typical LHC experiment resulting from the production and
subsequent annihilation of the lightest matter states in FSUSY, a pair
of right-handed f-squarks. The annihilation into a pair of hard
f-gluons and their fragmentation into glueballs  would provide
the first signal of FSUSY that is associated with the existence of a
hidden glue sector, a critical component of the theory.  
Our results show that it is likely that there will be more than one displaced
vertex per side of the event. Also, depending on the details of the
theory and most dramatically on the glueball mass, the distribution
of these vertices can fall almost anywhere in a typical LHC detector.
In Figures~\ref{fig:6} and \ref{fig:7} we exemplified this point by
choosing three values of the glueball mass, and various values for the
right-handed squark masses that define the hard f-gluon energy. We see
that for the lighter glueball mass preferred by our estimates,
i.e. $M_{\tilde{G}} =15~$GeV, most of the displaced vertices occur in
the outer elements of the detector, such as the muon systems or even
beyond, as seen in Figure~\ref{fig:6}. On the other hand, since the computation of the glueball mass
is rather uncertain, if we allow for somewhat larger values of
$M_{\tilde{G}}$, the situation changes dramatically and most of the
events are contained in the inner segments of the detector. 
It will be important then to develop searches capable of observing
multi-vertex events in virtually any part of the detector, and
possibly in more than one element at the same time. 

Since the cross sections producing these events are below a $1 fb$
(see Figure~\ref{fig:1}) large luminosities will be necessary to see
this physics above background. This study is a step in the direction
of modeling the ensuing multi-displaced vertex signals.  Next, it
will be necessary to fix further the parameters of the glueball fragmentation
function by  ``measuring'' the average multiplicity at one value of
the energy in a Monte Carlo shower simulation. This would put our
results for $D(x,s)$ on a better footing. Finally, a better modeling
of the glueballs resulting from soft f-gluon radiation in the event
after f-squark annihilation, would be helpful before a detailed
simulation for high luminosities is undertaken.

\acknowledgments
The authors thank Z.~Chacko, D.~Curtin and R.~Sundrum for helpful
discussions. 
They acknowledge 
the support of FAPESP, CNPq and CAPES.

\bibliography{fglueball}

\begin{thebibliography}{10}

\bibitem{Aad:2012tfa}
Georges Aad et~al.
\newblock {Observation of a new particle in the search for the Standard Model
  Higgs boson with the ATLAS detector at the LHC}.
\newblock {\em Phys. Lett.}, B716:1--29, 2012.

\bibitem{Chatrchyan:2012xdj}
Serguei Chatrchyan et~al.
\newblock {Observation of a new boson at a mass of 125 GeV with the CMS
  experiment at the LHC}.
\newblock {\em Phys. Lett.}, B716:30--61, 2012.

\bibitem{Martin:1997ns}
Stephen~P. Martin.
\newblock {A Supersymmetry primer}.
\newblock pages 1--98, 1997.
\newblock [Adv. Ser. Direct. High Energy Phys.18,1(1998)].

\bibitem{Aitchison:2005cf}
Ian J.~R. Aitchison.
\newblock {Supersymmetry and the MSSM: An Elementary introduction}.
\newblock 2005.

\bibitem{Bellazzini:2014yua}
Brando Bellazzini, Csaba Csaki, and Javi Serra.
\newblock {Composite Higgses}.
\newblock {\em Eur. Phys. J.}, C74(5):2766, 2014.

\bibitem{Panico:2015jxa}
Giuliano Panico and Andrea Wulzer.
\newblock {The Composite Nambu-Goldstone Higgs}.
\newblock {\em Lect. Notes Phys.}, 913:pp.1--316, 2016.

\bibitem{Chacko:2005pe}
Z.~Chacko, Hock-Seng Goh, and Roni Harnik.
\newblock {The Twin Higgs: Natural electroweak breaking from mirror symmetry}.
\newblock {\em Phys. Rev. Lett.}, 96:231802, 2006.

\bibitem{Cai:2008au}
Haiying Cai, Hsin-Chia Cheng, and John Terning.
\newblock {A Quirky Little Higgs Model}.
\newblock {\em JHEP}, 05:045, 2009.

\bibitem{Burdman:2006tz}
Gustavo Burdman, Z.~Chacko, Hock-Seng Goh, and Roni Harnik.
\newblock {Folded supersymmetry and the LEP paradox}.
\newblock {\em JHEP}, 02:009, 2007.

\bibitem{Craig:2014aea}
Nathaniel Craig, Simon Knapen, and Pietro Longhi.
\newblock {Neutral Naturalness from Orbifold Higgs Models}.
\newblock {\em Phys. Rev. Lett.}, 114(6):061803, 2015.

\bibitem{Craig:2014roa}
Nathaniel Craig, Simon Knapen, and Pietro Longhi.
\newblock {The Orbifold Higgs}.
\newblock {\em JHEP}, 03:106, 2015.

\bibitem{Craig:2015pha}
Nathaniel Craig, Andrey Katz, Matt Strassler, and Raman Sundrum.
\newblock {Naturalness in the Dark at the LHC}.
\newblock {\em JHEP}, 07:105, 2015.

\bibitem{Barbieri:2015lqa}
Riccardo Barbieri, Davide Greco, Riccardo Rattazzi, and Andrea Wulzer.
\newblock {The Composite Twin Higgs scenario}.
\newblock {\em JHEP}, 08:161, 2015.

\bibitem{Low:2015nqa}
Matthew Low, Andrea Tesi, and Lian-Tao Wang.
\newblock {Twin Higgs mechanism and a composite Higgs boson}.
\newblock {\em Phys. Rev.}, D91:095012, 2015.

\bibitem{Gherghetta:2016bcc}
Tony Gherghetta, Minh Nguyen, and Zachary Thomas.
\newblock {Neutral Naturalness with Bifundamental Gluinos}.
\newblock {\em Phys. Rev.}, D94(11):115008, 2016.

\bibitem{Craig:2016kue}
Nathaniel Craig, Simon Knapen, Pietro Longhi, and Matthew Strassler.
\newblock {The Vector-like Twin Higgs}.
\newblock {\em JHEP}, 07:002, 2016.

\bibitem{Serra:2017poj}
Javi Serra and Riccardo Torre.
\newblock {Neutral naturalness from the brother-Higgs model}.
\newblock {\em Phys. Rev.}, D97(3):035017, 2018.

\bibitem{Cohen:2018mgv}
Timothy Cohen, Nathaniel Craig, Gian~F. Giudice, and Matthew Mccullough.
\newblock {The Hyperbolic Higgs}.
\newblock {\em JHEP}, 05:091, 2018.

\bibitem{Cheng:2018gvu}
Hsin-Chia Cheng, Lingfeng Li, Ennio Salvioni, and Christopher~B. Verhaaren.
\newblock {Singlet Scalar Top Partners from Accidental Supersymmetry}.
\newblock {\em JHEP}, 05:057, 2018.

\bibitem{Curtin:2015fna}
David Curtin and Christopher~B. Verhaaren.
\newblock {Discovering Uncolored Naturalness in Exotic Higgs Decays}.
\newblock {\em JHEP}, 12:072, 2015.

\bibitem{Burdman:2008ek}
Gustavo Burdman, Z.~Chacko, Hock-Seng Goh, Roni Harnik, and Christopher~A.
  Krenke.
\newblock {The Quirky Collider Signals of Folded Supersymmetry}.
\newblock {\em Phys. Rev.}, D78:075028, 2008.

\bibitem{Chacko:2015fbc}
Zackaria Chacko, David Curtin, and Christopher~B. Verhaaren.
\newblock {A Quirky Probe of Neutral Naturalness}.
\newblock {\em Phys. Rev.}, D94(1):011504, 2016.

\bibitem{Strassler:2006im}
Matthew~J. Strassler and Kathryn~M. Zurek.
\newblock {Echoes of a hidden valley at hadron colliders}.
\newblock {\em Phys. Lett.}, B651:374--379, 2007.

\bibitem{Strassler:2006ri}
Matthew~J. Strassler and Kathryn~M. Zurek.
\newblock {Discovering the Higgs through highly-displaced vertices}.
\newblock {\em Phys. Lett.}, B661:263--267, 2008.

\bibitem{Han:2007ae}
Tao Han, Zongguo Si, Kathryn~M. Zurek, and Matthew~J. Strassler.
\newblock {Phenomenology of hidden valleys at hadron colliders}.
\newblock {\em JHEP}, 07:008, 2008.

\bibitem{Juknevich:2009ji}
Jose~E. Juknevich, Dmitry Melnikov, and Matthew~J. Strassler.
\newblock {A Pure-Glue Hidden Valley I. States and Decays}.
\newblock {\em JHEP}, 07:055, 2009.

\bibitem{Juknevich:2010rhj}
Jose~E. Juknevich.
\newblock {\em {Phenomenology of pure-gauge hidden valleys at Hadron
  colliders}}.
\newblock PhD thesis, Rutgers U., Piscataway, 2010.

\bibitem{Curtin:2018mvb}
David Curtin et~al.
\newblock {Long-Lived Particles at the Energy Frontier: The MATHUSLA Physics
  Case}.
\newblock 2018.

\bibitem{Sopczak:2017ezw}
Andre Sopczak.
\newblock {SUSY (ATLAS)}.
\newblock In {\em {6th International Conference on New Frontiers in Physics
  (ICNFP 2017) Kolymbari, Crete, Greece, August 17-26, 2017}}, 2017.

\bibitem{Lacroixon:2018lcy}
F.~Lacroixon.
\newblock {SUSY searches with the CMS detector}.
\newblock {\em Nuovo Cim.}, C40(5):189, 2017.

\bibitem{Kang:2008ea}
Junhai Kang and Markus~A. Luty.
\newblock {Macroscopic Strings and 'Quirks' at Colliders}.
\newblock {\em JHEP}, 11:065, 2009.

\bibitem{Harnik:2011mv}
Roni Harnik, Graham~D. Kribs, and Adam Martin.
\newblock {Quirks at the Tevatron and Beyond}.
\newblock {\em Phys. Rev.}, D84:035029, 2011.

\bibitem{Fok:2011yc}
R.~Fok and Graham~D. Kribs.
\newblock {Chiral Quirkonium Decays}.
\newblock {\em Phys. Rev.}, D84:035001, 2011.

\bibitem{Burdman:2014zta}
Gustavo Burdman, Zackaria Chacko, Roni Harnik, Leonardo de~Lima, and
  Christopher~B. Verhaaren.
\newblock {Colorless Top Partners, a 125 GeV Higgs, and the Limits on
  Naturalness}.
\newblock {\em Phys. Rev.}, D91(5):055007, 2015.

\bibitem{Delgado:1998qr}
A.~Delgado, A.~Pomarol, and M.~Quiros.
\newblock {Supersymmetry and electroweak breaking from extra dimensions at the
  TeV scale}.
\newblock {\em Phys. Rev.}, D60:095008, 1999.

\bibitem{Aad:2014fha}
Georges Aad et~al.
\newblock {Search for new resonances in $W\gamma$ and $Z\gamma$ final states in
  $pp$ collisions at $\sqrt s=8$ TeV with the ATLAS detector}.
\newblock {\em Phys. Lett.}, B738:428--447, 2014.

\bibitem{Morningstar:1999rf}
Colin~J. Morningstar and Mike~J. Peardon.
\newblock {The Glueball spectrum from an anisotropic lattice study}.
\newblock {\em Phys. Rev.}, D60:034509, 1999.

\bibitem{Chen:2005mg}
Y.~Chen et~al.
\newblock {Glueball spectrum and matrix elements on anisotropic lattices}.
\newblock {\em Phys. Rev.}, D73:014516, 2006.

\bibitem{Gockeler:2005rv}
M.~Gockeler, R.~Horsley, A.~C. Irving, D.~Pleiter, P.~E.~L. Rakow,
  G.~Schierholz, and H.~Stuben.
\newblock {A Determination of the Lambda parameter from full lattice QCD}.
\newblock {\em Phys. Rev.}, D73:014513, 2006.

\bibitem{Furmanski:1980cm}
W.~Furmanski and R.~Petronzio.
\newblock {Singlet Parton Densities Beyond Leading Order}.
\newblock {\em Phys. Lett.}, 97B:437--442, 1980.

\bibitem{Ellis:1991qj}
R.~Keith Ellis, W.~James Stirling, and B.~R. Webber.
\newblock {QCD and collider physics}.
\newblock {\em Camb. Monogr. Part. Phys. Nucl. Phys. Cosmol.}, 8:1--435, 1996.

\bibitem{Wakely:1991eu}
A.~B. Wakely and C.~E. Carlson.
\newblock {Pseudoscalar glueball wave functions from QCD sum rules}.
\newblock {\em Phys. Rev.}, D45:338--343, 1992.

\bibitem{Wakely:1991ej}
A.~B. Wakely and C.~E. Carlson.
\newblock {Two photon production of pseudoscalar glueballs}.
\newblock {\em Phys. Rev.}, D45:1796--1799, 1992.

\bibitem{Roy:1999bu}
Probir Roy and K.~Sridhar.
\newblock {A New proposal for glueball exploration in hard gluon
  fragmentation}.
\newblock {\em JHEP}, 07:013, 1999.

\bibitem{Chiappetta:1992uh}
P.~Chiappetta, Mario Greco, J.~P. Guillet, S.~Rolli, and M.~Werlen.
\newblock {Next-to-leading order determination of pion fragmentation
  functions}.
\newblock {\em Nucl. Phys.}, B412:3--38, 1994.

\bibitem{Patrignani:2016xqp}
C.~Patrignani et~al.
\newblock {Review of Particle Physics}.
\newblock {\em Chin. Phys.}, C40(10):100001, 2016.

\bibitem{Meyer:2008tr}
Harvey~B. Meyer.
\newblock {Glueball matrix elements: A Lattice calculation and applications}.
\newblock {\em JHEP}, 01:071, 2009.

\bibitem{Aad:2015uaa}
Georges Aad et~al.
\newblock {Search for long-lived, weakly interacting particles that decay to
  displaced hadronic jets in proton-proton collisions at $\sqrt{s}=8$ TeV with
  the ATLAS detector}.
\newblock {\em Phys. Rev.}, D92(1):012010, 2015.

\bibitem{Sirunyan:2017jdo}
Albert~M Sirunyan et~al.
\newblock {Search for new long-lived particles at $\sqrt{s} =$ 13 TeV}.
\newblock {\em Phys. Lett.}, B780:432--454, 2018.

\end{thebibliography}
\bibliographystyle{unsrt}

\end{document}